\documentclass[aps,twocolumn,prd]{revtex4-2}

\usepackage{bm,graphicx,textcomp,amssymb,amsmath,dcolumn,color}
\usepackage{tikz,pgfplots}
\usetikzlibrary{decorations.markings,snakes} 
\usetikzlibrary{arrows,calc,decorations.markings,math,arrows.meta}

\newcommand{\ket}[1]{\left|#1\right>}
\newcommand{\bra}[1]{\left<#1\right|}
\newcommand{\braket}[2]{\left<#1|#2\right>}

\newcommand{\f}[1]{\mbox{\boldmath$#1$}}

\newcommand{\bea}{\begin{eqnarray}}
	\newcommand{\ea}{\end{eqnarray}}
\newcommand{\eea}{\end{eqnarray}}
\newcommand{\ord}{\,{\cal O}}

\begin{document}

\title{Stimulated emission or absorption of gravitons by light} 

\author{Ralf Sch\"utzhold}

\email{r.schuetzhold@hzdr.de}

\affiliation{Helmholtz-Zentrum Dresden-Rossendorf, 
Bautzner Landstra{\ss}e 400, 01328 Dresden, Germany,}

\affiliation{Institut f\"ur Theoretische Physik, 
Technische Universit\"at Dresden, 01062 Dresden, Germany,}

\date{\today}

\begin{abstract}
We study the exchange of energy between gravitational and electromagnetic waves
in a Sagnac type geometry, in analogy to an ``optical Weber bar.''
In the presence of a gravitational wave (such as the ones measured by LIGO), 
we find that it should be possible to observe signatures of stimulated emission
or absorption of gravitons with present day technology.
Apart from marking the transition from passively observing to actively manipulating
such a natural phenomenon, this could also be used as a complementary 
detection scheme.
Non-classical photon states may improve the sensitivity and might even allow us
to test certain quantum aspects of the gravitational field.
\end{abstract}

\maketitle

\section{Introduction}

In the history of electrodynamics, an important step was Franklin's pioneering 
(though extremely dangerous) kite experiment where static electricity was 
collected from the air by flying a kite into or close to thunder clouds.
On the one hand, this experiment showed that lightning and electricity as known 
from laboratory experiments (e.g., with Leiden jars or by rubbing amber) are  
basically of the same nature and thereby made a significant contribution to 
unifying these phenomena -- eventually leading to our modern understanding of 
electrodynamics and the standard model.
On the other hand, the kite experiment marked the transition from passively 
observing a natural phenomenon such as lightning to actively manipulating it 
-- and thereby paved the way for many modern technological developments, 
from lightning rods to power plants etc. 

Due to the weakness of the gravitational interaction (in laboratory scale experiments)
as determined by Newton's constant
$G_{\rm N}\approx6.7\times10^{-11}\,\rm m^3\,kg^{-1}\,s^{-2}$,
we are now in a somewhat similar situation regarding gravitational waves.
They have been predicted by Einstein around a century ago 
\cite{Einstein-1916,Einstein-1918}. 
%
%
However, it took more than half a century before indirect evidence for them 
has been observed in the Hulse-Taylor binary pulsar 
\cite{Hulse:1974eb,Taylor:1982zz} whose energy loss over time due to the 
emission of gravitational waves agrees very well with the predictions from 
general relativity. 
Recently, the direct detection of gravitational waves on earth has been achieved 
at LIGO \cite{LIGO-paper,LIGO-www}. 
Both accomplishments mark important breakthroughs and have been awarded with the 
Nobel prizes in physics in 1993 and 2017, respectively. 

In the following, we shall make the assumption (which is quite natural but not 
verified experimentally yet) that gravitational waves are, 
at least in the weak-field regime, analogous to electromagnetic waves in the sense 
that their energy is quantized in terms of excitation quanta $\hbar\omega$
(i.e., gravitons) where $\omega$ is the frequency of the gravitational wave.
Possible consequences of departures from this assumption will be discussed below.
Then, in order to facilitate the transition from passively observing a natural 
phenomenon such as gravitational waves to actively manipulating it, let us ask 
the following question: 
Can one design an experiment where at least one graviton with energy 
$\hbar\omega$ is emitted (or absorbed) in a verifiable manner? 
As a first approach to this question, let us take the well-known quadrupole formula 
describing the power emitted by gravitational radiation
(in analogy to the dipole formula in electromagnetism) 
\bea
\label{quadrupole}
P=\frac{G_{\rm N}}{45c^5}\sum_{ij}\left(\stackrel{...}{Q}_{ij}\right)^2 
\,,
\ea
where $Q_{ij}$ are the quadrupole moments of the dynamical mass distribution
\cite{Wald:1984rg}.
In terms of its characteristic length $L$ and mass $m$, 
they scale as $Q_{ij}=\ord(mL^2)$.
Thus the total emitted power goes as 
$P=\ord(\omega^6m^2L^4G_{\rm N}/c^5)$ where $\omega$ is the oscillation 
frequency (i.e., the frequency of the emitted gravitational waves).
Together with Newton's constant $G_{\rm N}$, the 
speed of light $c\approx3\times10^8\,\rm m/s$ 
suppresses the pre-factor 
in the quadrupole formula~\eqref{quadrupole} 
by more than fifty orders of magnitude when expressed in terms of SI units,
see also \cite{Wald:1984rg}. 

Thus, even after comparison to the small Planck constant 
$\hbar\approx10^{-34}\,\rm Js$, we find that it will be extremely hard to emit 
one gravitational excitation quantum (graviton) with energy $\hbar\omega$ 
using everyday values of $m$, $L$ and $\omega$ in the kilogram, 
meter and seconds (Hertz) regime \cite{footnote-kHz}.
In order to cast the result into a dimensionless form, let us introduce  
the number $N$ of gravitons emitted per oscillation period 
$N=\ord(P/[\hbar\omega^2])$, the characteristic velocity scale 
$v=\omega L$ and the Planck mass 
$m_{\rm P}=\sqrt{\hbar c/G_{\rm N}}\approx22~\mu\rm g$.
Then we find that $N$ scales as $m^2/m_{\rm P}^2$ multiplied by $v^4/c^4$ 
showing a strong suppression for slow velocities, i.e., 
in the non-relativistic regime.  

These considerations suggest using light \cite{footnote-charged}.
Since stationary CW laser beams do not emit gravitational waves
\cite{footnote-stationary}, we consider laser pulses, 
see also \cite{Spengler:2021rlg}.
As an extreme example, let us take the Mega-Joule pulses 
at the National Ignition Facility (NIF) \cite{NIF}. 
Still, they only correspond to a mass of order $10^{-11}\,\rm kg$,
i.e., well below the Planck mass.
As a result, already a rough order-of-magnitude estimate obtained by 
combining this mass of order $10^{-11}\,\rm kg$ (squared) with 
Newton's constant $G_{\rm N}$ in comparison to 
$\hbar c\approx3\times10^{-26}\,\rm J\,m$ shows 
that it is still very hard to emit a single graviton in this way. 

Hence, we follow a different route here and consider the stimulated emission 
of gravitons instead of their creation by the quadrupole formula~\eqref{quadrupole}, 
see also \cite{Boughn:2006st,Rothman:2006fp,Giovannini:2019ehc}
(though in a different context). 
To this end, we consider light pulses propagating within a pre-existing gravitational 
wave (as the ones measured by LIGO, for example) and determine the transfer of 
energy between the gravitational and the electromagnetic field, 
cf.~\cite{Grafe:2023ngy}. 
Finding an energy transfer of $\hbar\omega$ or more is then interpreted as a 
smoking gun for the emission or absorption of gravitons by light. 
Note that this scheme displays some similarities to resonant mass antennas such as 
Weber bars \cite{Weber-1967,Weber-1968,Weber-1969} but since we are using a highly 
excited state (light pulse), we may not only absorb but also emit gravitational 
radiation. 

\section{Gravitational waves}

For simplicity, let us consider linearly polarized gravitational waves propagating in 
$z$-direction, but our results can be generalized to other wave forms  in a 
straightforward way.
In a suitable coordinate system, the metric reads $(\hbar=c=\varepsilon_0=\mu_0=1)$
\bea
\label{metric}
ds^2=dt^2-[1+h]dx^2-[1-h]dy^2-dz^2
\,,
\ea
where $h(t,z)=h(t-z)$ is the amplitude of the gravitational wave.
Since this quantity is extremely small, such as $h=\ord(10^{-22})$, 
we neglect second and higher orders in the following. 
Thus, the metric determinant simplifies to $\sqrt{-g}=1+\ord(h^2)$.
Furthermore, in view of the long wavelength of the gravitational waves and 
the fact that we consider light pulses propagating in the $x,y$-plane, 
we may neglect the spatial dependence $h(t,z)\approx h(t)$. 

In these coordinates~\eqref{metric}, the Christoffel symbols corresponding 
to Newton's gravitational acceleration vanish $\Gamma^i_{00}=0$ and thus massive 
objects such as the mirrors used to reflect the light pulses stay at rest.
However, since the $x$ and $y$ coordinates are re-scaled differently by the 
gravitational wave, it could affect the angle under which the light pulses are 
reflected. 
In principle, this angular deflection of order $\ord(h)$ could also be used to 
detect gravitational waves. 
However, since laser beams/pulses with a well-defined propagation direction must 
have a sufficiently large width (of many wavelengths), the impact of this tiny 
deflection angle can be neglected here. 

\section{Electromagnetic waves}

Now let us consider light pulses propagating in the background~\eqref{metric}.
In order to maximize the effect, we consider light polarized in $z$-direction 
$\f{A}(t,\f{r})=A_z(t,x,y)\f{e}_z$ but again our analysis can easily be generalized. 
Note that this form $\f{A}(t,\f{r})=A_z(t,x,y)\f{e}_z$
automatically satisfies the generally relativistic 
Lorenz gauge condition $\nabla_\mu A^\mu=0$. 
The contraction $F_{\mu\nu}F^{\mu\nu}$ of the field strength tensor
$F_{\mu\nu}=\partial_\mu A_\nu-\partial_\nu A_\mu$ gives the Lagrangian density 
\bea
\label{Lagrangian}
{\cal L}
=
\frac12\left[
(\partial_tA_z)^2
-[1-h](\partial_xA_z)^2
-[1+h](\partial_yA_z)^2
\right] 
\,.
\ea
Field quantization yields the interaction Hamiltonian 
\bea
\hat H_{\rm int}=h\int d^3r
\left[(\partial_y\hat A_z)^2-(\partial_x\hat A_z)^2\right]
\,,
\ea
which is determined by the magnetic fields $\hat B_x^2-\hat B_y^2$ 
and describes the coupling between the electromagnetic field and the 
gravitational wave. 

Now we may study the energy transferred between these two. 
To this end, we employ the Heisenberg picture with 
$d\hat H/dt=(\partial\hat H/\partial t)_{\rm expl}$ 
where the explicit time dependence stems from the gravitational
wave, i.e., $h(t)$.
Taking expectation values then yields the energy transfer 
\bea
\label{transfer}
\frac{d\langle\hat H\rangle}{dt}
=
\dot h\int d^3r
\left\langle(\partial_y\hat A_z)^2-(\partial_x\hat A_z)^2\right\rangle 
\,,
\ea
where the divergent vacuum contributions $\bra{0}(\partial_y\hat A_z)^2\ket{0}$
and $\bra{0}(\partial_x\hat A_z)^2\ket{0}$ cancel each other such that we may 
use renormalized (e.g., normal ordered) values.
The integral on the right-hand side of Eq.~\eqref{transfer} is the difference
between the total energies of the light pulse in the magnetic field components
in $x$ and $y$ direction. 
Thus we find a rigorous bound $|\dot E|\leq|\dot h|E$ for the energy transfer 
$\dot E=d\langle\hat H\rangle/dt$ in terms of the total energy $E$ 
of the laser pulse \cite{footnote-energy-inequalities}. 
In practise, however, it is very hard to saturate this bound since the light 
energy oscillates rapidly between the electric component 
$\langle(\partial_t\hat A_z)^2\rangle_{\rm ren}$
and the magnetic components 
$\langle(\partial_x\hat A_z)^2\rangle_{\rm ren}$ or 
$\langle(\partial_y\hat A_z)^2\rangle_{\rm ren}$.
Thus we have $|\dot E|\leq|\dot h|E/2$ on average. 

In order to maximize energy transfer, one could imagine the following scenario,
see also Fig.~\ref{figure}. 
As long as $\dot h>0$, we have a light pulse propagating in $x$-direction, 
and then -- after reflection by a mirror -- it propagates in $y$-direction 
as long as $\dot h<0$, and so on. 
In this case, we have $\dot E<0$ and thus the emission of gravitons 
(in view of energy conservation). 
The opposite case ($x$-direction for $\dot h<0$ and $y$-direction for $\dot h>0$)
yields $\dot E>0$ and thus the absorption of gravitons.

\section{Wave packets}

Let us consider the sequence described above in terms of the wave packets associated to 
the light pulses. 
The wave equation obtained from the Lagrangian~\eqref{Lagrangian} reads 
\bea
\label{wave-equation} 
\left(
\partial_t^2
-[1-h]\partial_x^2
-[1+h]\partial_y^2
\right)A_z=0
\,.
\ea
Since the frequency $\Omega=\ord(10^{15}\,\rm Hz)$ of the electromagnetic waves 
corresponding to visible or near infra-red photons with energies in the eV regime
is much larger than frequency $\omega$ of the gravitational wave 
(e.g., in the kHz or the Hz range), we may use the WKB approximation. 
In the coordinates~\eqref{metric}, the wave-numbers $K_x$ and $K_y$ are conserved
(apart from the reflection at the mirrors), but the frequencies $\Omega$ change 
according to the dispersion relation 
\bea
\Omega^2=[1-h]K_x^2+[1+h]K_y^2
\,.
\ea
For propagation in either $x$-direction or $y$-direction, the 
energy of each photon thus changes with $\Delta\Omega=\pm h\Omega/2$.
During the reflections at the static mirrors (occurring when $\dot h=0$), 
the frequencies $\Omega$ do not change.
Hence, by altering the directions as described above, one can transform the 
momentary changes $\Delta\Omega=\pm h\Omega/2$ into a lasting shift in frequency. 
Since the total number of photons does not change in this process, we get 
an energy shift of $\Delta E=\pm h E/2$ for each half-period of the gravitational
wave, i.e., in between two reflections at the mirrors 
(consistent with the results above). 

Besides the total energy of the wave-packets (on the classical level), 
let us also consider its shape and amplitude. 
Since the values of the wave-numbers are conserved during free propagation 
(i.e., in between two reflections at the mirrors), the shape of the 
wave packet does not change in terms of the coordinates~\eqref{metric}.
However, due to the reflections at the mirrors 
(where the $x$ and $y$ length scales are modified by gravitational wave) 
we may get a lasting deformation of the wave packets, i.e., light pulses.
Similar to the deflection angle discussed above, these deformations could also 
be used to detect gravitational waves, at least in principle. 
However, since these deformations are very small $\ord(h)$, we may neglect them 
in the following and focus on the energy shift. 
Even after the interaction with the gravitational wave, the energy shift induces 
a phase difference which grows with the time elapsed and thus can be amplified 
-- while the deformation would not be amplified in the same way. 

Finally, for fixed $K_x$ and $K_y$ (i.e., in between two reflections at the mirrors),
the wave equation~\eqref{wave-equation} simplifies to the ordinary differential 
equation $\ddot A_z+\Omega^2(t)A_z=0$ with the conserved Wronskian 
$W=A_z^*\dot A_z-\dot A_z^*A_z$ which yields $W\approx-2i\Omega|A_z^2|$
in the WKB approximation. 
Thus, the amplitude of $A_z$ changes with $1/\sqrt{\Omega}$. 
Since the total energy $E$ of the pulse scales with $(\Omega A_z)^2$ 
and the volume in terms of the coordinates~\eqref{metric} does not change 
during free propagation, we find that $E$ changes proportional to 
$\Omega$ (i.e., the energy of each photon) as expected.  
Again, the reflections at the static mirrors (occurring when $\dot h=0$) 
do not change the total energy -- but they transform the instantaneous changes 
into a lasting energy shift. 

\section{Experimental parameters} 

\begin{figure}[t]
\begin{tikzpicture}[]
\draw[thick, black, dotted] (-0.3,-0.3) -- (0.3,0.3);
\draw[thick, red] (-0.5,0) -- (3,0);
\draw[thick, red] (0,0) -- (0,3);
\draw[thick, red] (0,3) -- (0.3,1);
\draw[thick, red] (0.3,1) -- (0.3,3.3);
\draw[thick, red] (0.3,3.3) -- (0.6,1.3);
\draw[thick, red] (0.6,1.3) -- (0.6,3.6);
\draw[thick, black] (0.4,3.4) -- (0.8,3.8);
\draw[thick, black] (3.4,0.4) -- (3.8,0.8);
\draw[thick, red] (3,0) -- (1,0.3);
\draw[thick, red] (1,0.3) -- (3.3,0.3);
\draw[thick, red] (3.3,0.3) -- (1.3,0.6);
\draw[thick, red] (1.3,0.6) -- (3.6,0.6);
\draw[thick, red] (3.6,0.6) -- (3.6,2.9);
\draw[thick, red] (3.6,2.9) -- (3.9,0.9);
\draw[thick, red] (3.9,0.9) -- (3.9,3.2);
\draw[thick, red] (3.9,3.2) -- (4.2,1.2);
\draw[thick, red] (4.2,1.2) -- (4.2,3.5);
\draw[thick, red] (0.6,3.6) -- (2.9,3.6);
\draw[thick, red] (2.9,3.6) -- (0.9,3.9);
\draw[thick, red] (0.9,3.9) -- (3.2,3.9);
\draw[thick, red] (3.2,3.9) -- (1.2,4.2);
\draw[thick, red] (1.2,4.2) -- (3.5,4.2);
\draw[thick, red] (4.2,3.5) -- (5.2,3.5);
\draw[thick, red] (5.2,3.5) -- (5.2,0);
\draw[thick, red] (5.2,0) -- (5.4,3.5);
\draw[thick, red] (5.4,3.5) -- (5.4,0);
\draw[thick, red] (5.4,0) -- (5.6,3.5);
\draw[thick, red] (5.6,3.5) -- (5.6,0);
\draw[thick, red] (5.6,0) -- (5.8,3.5);
\draw[thick, red] (5.8,3.5) -- (5.8,0);
\draw[thick, red] (3.5,4.2) -- (6.4,4.2);
\draw[thick, red] (6.4,4.2) -- (6.4,0.7);
\draw[thick, red] (6.4,0.7) -- (6.6,4.2);
\draw[thick, red] (6.6,4.2) -- (6.6,0.7);
\draw[thick, red] (6.6,0.7) -- (6.8,4.2);
\draw[thick, red] (6.8,4.2) -- (6.8,0.7);
\draw[thick, red] (6.8,0.7) -- (7,4.2);
\draw[thick, red] (7,4.2) -- (7,-0.5);
\draw[thick, red] (5.8,0) -- (7.5,0);
\draw[thick, black, dotted] (7.3,-0.3) -- (6.7,0.3);
\end{tikzpicture}
\caption{Sketch (not to scale) of a possible geometry. 
The initial laser pulse is split up by a half silvered mirror 
(dotted black line on bottom left) into two pulses (red lines) 
first propagating in $x$ and $y$ direction, respectively. 
After half a period (ideally at $\dot h=0$), these pulses are reflected 
by the $45^\circ$ mirrors (solid black lines) in order to 
propagate in the respective other directions. 
After traversing this Sagnac type geometry on the left-hand side
and thereby gaining or loosing energy, 
the light pulses are sent through further optical paths of equal length on the 
right-hand side in order to accumulate a large enough phase difference. 
Finally, they are brought to interference at another 
half silvered mirror (dotted black line on bottom right). 
The optical paths are elongated by retro-reflections and the mirrors 
for doing that and for guiding the pulses are not shown for simplicity.}
\label{figure}
\end{figure}
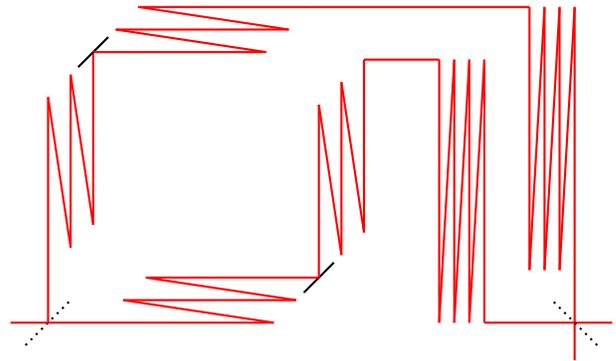

Let us study the experimental feasibility of the above scheme by inserting typical 
example values for the parameters. 
Assuming a gravitational wave with a frequency in the kHz regime corresponds to a 
propagation length of a few hundred km during one half-period. 
As in LIGO, this length can be folded into a smaller length scale by 
retro-reflecting mirrors \cite{footnote-retro}, see Fig.~\ref{figure}. 
Then, after a propagation time of order ms, the light pulses hit the 
$45^\circ$ mirrors which change their direction from 
$\f{K}_{\rm in}=K_x\f{e}_x$ to $\f{K}_{\rm out}=K_y\f{e}_y$
or vice versa. 
Ideally, this should happen when $\dot h=0$ such that the sign changes of 
$h(t)$ and the integrand in Eq.~\eqref{transfer} cancel each other. 
Depending on how monochromatic the gravitational wave is, one could repeat this 
procedure for a several half-cycles in order to obtain a lasting energy shift of 
several $hE$ which should then equal or exceed $\hbar\omega$.  

In this scheme, the light pulses must fit into one half-period of the 
gravitational wave, such that the pulse duration is well below ms and thus 
the frequency uncertainty well above kHz. 
Thus, the idea is to transfer the small energy shift $\Delta E\geq\hbar\omega$ 
into a phase shift $\Delta\varphi$ which can be measured via interference. 
To this end, the initial pulse could be spilt up via a half silvered mirror 
(non-polarizing beam splitter) at $45^\circ$ into two equal pulses, 
one first propagating in $x$-direction and the other one first propagating 
in $y$-direction, see Fig.~\ref{figure}. 
In this way, these two pulses would acquire opposite energy transfers. 

So far, the set-up is similar to (half of) a Sagnac interferometer, 
but as an important 
difference, one does not let the two light pulses interfere at this stage.
Instead, they are both sent through another optical path length 
(which can be basically the same for both pulses) during which their 
tiny and opposite energy shifts $\pm\Delta E$ generate a small phase 
difference $\Delta\varphi$. 
This phase accumulation period would be after the gravitational wave passed by 
and thus can be much longer than a period of the gravitational wave.
Actually, this fact could be an important advantage in comparison to LIGO, 
where the effective optical path length $\ord(10^3\,\rm km)$ 
is limited by the period of the gravitational wave such that 
one has a few hundred reflections at the mirrors before interference.
Of course, this advantage does also come along with a drawback, 
since LIGO can measure the full time-dependent amplitude $h(t)$ 
of the gravitational wave, while the scheme here focuses on the final 
energy shift. 

For photons with energies in the eV regime (visible or near infra-red light), 
a laser pulse with a moderate energy in the mJ regime contains $N=\ord(10^{16})$ 
photons.
In view of their frequency $\Omega=\ord(10^{15}\,\rm Hz)$, we see that 
gravitational waves with $h=\ord(10^{-22})$ or even weaker should lead to 
the stimulated emission (or absorption) of many gravitons with 
$\omega=\ord(\rm kHz)$, cf.~\cite{Lieu:2017lzh}. 
Assuming a classical (coherent) pulse, the usual Poisson limit 
$\Delta\varphi\propto1/\sqrt{N}$ yields the achievable phase accuracy 
$\Delta\varphi=\ord(10^{-8})$ for interference measurements. 
Non-classical photon states will be discussed below. 

Another enhancement factor $\ord(10^9)$ is the large ratio of length scales: 
$\ord(\rm km)$ arm length versus $\ord(\mu\rm m)$ wavelength.
These two enhancement mechanisms are basically the same as in LIGO. 
As a difference to LIGO, the lasting energy shift $\Delta E$ allows longer  
phase accumulation times. 
For example, an effective optical path length of $\ord(10^6\,\rm km)$, e.g., 
by assuming $\ord(10^6)$ reflections (instead of the few hundred reflections at LIGO), 
would yield a total enhancement factor of $\ord(10^{23})$ which looks very promising 
for amplitudes $h=\ord(10^{-22})$. 

Form another perspective, each photon acquires a lasting frequency 
shift of $\pm\Delta\Omega=\ord(h\Omega)$ which gives $\ord(10^{-7}\,\rm Hz)$.
After a phase accumulation time of a few seconds corresponding to a path length 
of $\ord(10^6\,\rm km)$, this translates into a phase 
shift of $\Delta\varphi=\ord(10^{-7})$ for each photon -- which can then be 
detected by using $N=\ord(10^{16})$ photons.

The above considerations assumed that the light pulses are perfectly timed with 
the gravitational waves such that the former hit the $45^\circ$ mirrors when 
$\dot h=0$. 
If this timing is not perfect, the effect is reduced accordingly. 
This drawback could be reduced by having several pulses emitted during 
one gravitational wave period -- ideally as coincidence measurement with LIGO. 
Note that the total average power of a few Watt is not overwhelming. 
Thus, going to the limit of overlapping pulses, one could also envision a CW laser 
with a permanent power in this range, where the interference pattern is continuously 
measured.  

\section{Non-classical photon states} 

It is well known that one can achieve sensitivities exceeding the Poisson limit 
$\Delta\varphi\propto1/\sqrt{N}$
by employing non-classical states such as squeezed states, see, e.g., 
\cite{Braginskii:1974zz,Unruh:1978gx,Unruh:1979de}.
Actually, this is being implemented at LIGO, cf.~\cite{LIGOScientific:2013pcc}. 
Since the energy transfer~\eqref{transfer} is bounded by the total energy of the 
light pulse (independent of its quantum state), a squeezed state would not be 
an advantage here -- except that it could have an energy variance which is different 
from a coherent state. 

However, a non-classical state can be advantageous for the accuracy of the 
phase measurement. 
To understand this point, let us consider the extreme case of a NOON state,
see, e.g., \cite{Dowling:2008pbf}. 
In contrast to a coherent (i.e., classical) state where all photons are in a 
superposition of the two interferometer arms, this NOON state describes a 
superposition where either {\em all} photons are in one arm (and none in the other)
or {\em all} photons are in the other arm 
$\ket{{\rm NOON}}=(\ket{N}\ket{0}+\ket{0}\ket{N})/\sqrt{2}$.
After interacting with the gravitational wave, the photons acquire opposite phases 
$(e^{+iN\Delta\varphi}\ket{N}\ket{0}+e^{-iN\Delta\varphi}\ket{0}\ket{N})/\sqrt{2}$ 
such that now the achievable phase sensitivity scales with the Heisenberg limit 
$\Delta\varphi=\ord(1/N)$ instead of the Poisson limit 
$\Delta\varphi=\ord(1/\sqrt{N})$. 
As a result, the required number $N$ of photons would be much smaller, 
but actually generating such a highly non-classical state is also much 
more challenging experimentally. 

\section{Quantum aspects of gravity}

So far, we have assumed that the laws of quantum theory apply to the gravitational 
field in basically the same way as to the electromagnetic field, for example. 
Now, let us scrutinize this assumption.
First, it should be stressed that measuring an energy shift of $\hbar\omega$ does 
{\em not} prove that the energy of gravitational waves is quantized in units of
$\hbar\omega$. 
On the other hand, detecting a gravitational wave at LIGO, for example, and {\em not}
finding the associated energy transfer in a set-up discussed here would indicate that
there is something going on we do not understand (e.g., that the above assumption is
wrong).

Furthermore, the set-up discussed here could allow us to test certain properties
of quantum superposition states of gravitational fields. 
Similar ideas have already been discussed for the Newtonian gravitational field:
If a sufficiently large mass is in a superposition state of two spatially well
separated positions, then its (static) gravitational field should also be in a 
quantum superposition, see, e.g.,
\cite{Baym:2009zu,Romero-Isart:2011yun,Mari:2015qva,Belenchia:2018szb}. 
Going one step further, this superposition state could indicate entanglement 
between the gravitational field and the matter degrees of freedom -- or even 
mediate entanglement between two different matter degrees of freedom, 
see, e.g., \cite{Altamirano:2016fas,Marletto:2017kzi}.

An analogous idea can be applied to the set-up considered here.
For example, let us take the NOON state discussed above for the photon field
where the light pulse in one arm (say, $\ket{N}\ket{0}$) would gain the 
energy $\Delta E$ while the light pulse in the other arm ($\ket{0}\ket{N}$)
would loose this energy. 
Then, unless one is willing to abandon energy conservation, this means that 
we get a superposition of quantum states including the gravitational wave, 
i.e., $\ket{{\rm NOON}}\ket{\bar E_{\rm grav}}$ transforms to a superposition of
the state 
$e^{+iN\Delta\varphi}\ket{N}\ket{0}\ket{\bar E_{\rm grav}-\Delta E}$
for one arm and the state 
$e^{-iN\Delta\varphi}\ket{0}\ket{N}\ket{\bar E_{\rm grav}+\Delta E}$
for the other arm, where $\bar E_{\rm grav}$ denotes the energy expectation 
value of the gravitational wave. 

However, this superposition 
does not necessarily imply strong entanglement between the photon field and
the gravitational field because the quantum states of the latter
$\ket{\bar E_{\rm grav}-\Delta E}$ and $\ket{\bar E_{\rm grav}+\Delta E}$
are not necessarily orthogonal.
For example, two coherent states $\ket{\alpha_1}$ and $\ket{\alpha_2}$
have a finite overlap
$|\braket{\alpha_1}{\alpha_2}|^2=\exp\{-|\alpha_1-\alpha_2|^2\}$
which can be near unity if the two states $\ket{\alpha_1}$ and $\ket{\alpha_2}$
just differ by an energy of one excitation quantum $\hbar\omega$ on top of 
strongly displaced (i.e., nearly classical) state with $|\alpha|\gg1$.
This would be very different for a Fock state $\ket{n}$, for example, where 
$\ket{n}$ and $\ket{n+1}$ are orthogonal for all $n$ and thus the overlap 
vanishes (i.e., one has maximum entanglement). 

This entanglement between the photon field and the gravitational field 
or the overlap between the states $\ket{\bar E_{\rm grav}-\Delta E}$ and
$\ket{\bar E_{\rm grav}+\Delta E}$ affects the visibility in interference
measurements of the phase difference $\Delta\varphi$. 
For an overlap of unity (i.e., no entanglement), one has full visibility 
while a vanishing overlap (i.e., maximum entanglement) results in the absence 
of any interference. 
Thus, by measurements on the photon field
(e.g., using the NOON states with variable delay times), 
one can in principle distinguish 
different quantum states of the gravitational field (in this mode), such as 
a coherent state or a Fock state or a thermal state. 

In a bigger picture, the above consideration is an example of gravitational
decoherence which has already been discussed in several works, see, e.g., 
\cite{Zych:2012ut,Pikovski:2013qwa,Blencowe:2012mp,Anastopoulos:2013zya,Bassi:2017szd}.
The arguments above are based on the assumption that the laws of quantum
theory apply to gravity in the same way as to electromagnetism, for example,
but it has also been proposed that one has to modify gravity and/or quantum
theory when combining them, see, e.g., 
\cite{Ghirardi:1985mt,Diosi:1989hlx,Penrose:1996cv}. 
In such a case, the predictions could be different (e.g., the decoherence
could be larger) and thus the set-up could also allow us to test these ideas, 
see also \cite{Kafri:2014zsa,Carney:2018ofe,footnote-models}. 

As another potentially interesting observable, one could measure the phase
fluctuations (for the pulsed mode of operation or the CW mode).
Since the final phase $\Delta\varphi(t)$ is given by a convolution of the
gravitational wave amplitude $h(t')$ with a time dependent kernel $k(t-t')$
which encodes the history (e.g., reflections) of the photons arriving at a
time $t$, these phase fluctuations $\langle(\Delta\hat\varphi)^2\rangle$
allow us to access the two-point function $\langle\hat h(t)\hat h(t')\rangle$
of the graviton field.
This quantity contains information 
(phase coherence versus thermal fluctuations etc.) 
about the quantum state of the graviton field.

\section{Conclusions and Outlook}

In order to facilitate the transition from passively observing a natural phenomenon 
such as a gravitational wave to actively manipulating it, we investigate the stimulated
emission or absorption of gravitons by light \cite{footnote-barbell}, 
in analogy to an ``optical Weber bar.''
An important difference to LIGO is the distinction between the interaction time 
(set by the period and pulse length of the gravitational wave) and the 
phase accumulation time. 
For LIGO, both are essentially the same, but in the set-up discussed here,
the latter is not limited by the gravitational wave but only by optical 
properties (such as Q factor) and thus could be much longer.  
This difference might become even more pronounced for gravitational waves 
of higher frequencies. 

Using non-classical photon states such as NOON states, energy conservation
demands that we create quantum superposition states of gravitational waves 
with different energies. 
In this way, interference experiments with variable delay times could even 
test certain quantum aspects of gravity, e.g., 
distinguish between different quantum states of the gravitational field, 
such as coherent states, squeezed states, Fock states, or thermal states. 

\acknowledgments 

Fruitful discussions with S.~Liberati, D.~Page and W.G.~Unruh 
are gratefully acknowledged. 



\begin{thebibliography}{99}
\bibitem{footnote-kHz} 
In view of the $\omega^6$ scaling of the emitted power~\eqref{quadrupole},
going to much higher frequencies could help, but imagining masses 
$m=\ord(\rm kg)$ oscillating with amplitudes $L=\ord(\rm m)$ at frequencies
in the kHz regime -- and detecting small energy shifts of $\hbar\omega$ -- 
shows that such an experiment is also extremely challenging.
\bibitem{footnote-charged}
Charged particles such as protons or electrons can also be accelerated to 
nearly the speed of light, but in this case the electromagnetic radiation 
would dominate any gravitational effects by far.
\bibitem{footnote-stationary}
Note that this applies to the standard coupling $h^{\mu\nu}T_{\mu\nu}$ 
to the energy-momentum tensor $T_{\mu\nu}$ in linearized general relativity.
%
For other (hypothetical) coupling terms between photons and gravitons, 
the situation might be different.
%
However, even in those cases, symmetries such as Lorentz invariance 
restrict such photon-graviton conversion processes strongly.
\bibitem{footnote-energy-inequalities} 
Even though the renormalized expectation values 
$\langle(\partial_x\hat A_z)^2\rangle_{\rm ren}$ or 
$\langle(\partial_y\hat A_z)^2\rangle_{\rm ren}$
could also become negative in some regions of space-time, the total 
volume integral of them must be non-negative.
\bibitem{footnote-retro} 
These mirrors, however, do not change the propagation direction from 
$\f{K}_{\rm in}=K_x\f{e}_x$ to $\f{K}_{\rm out}=K_y\f{e}_y$
but from $\f{K}_{\rm in}=K_x\f{e}_x$ to $\f{K}_{\rm out}=-K_x\f{e}_x$
or from $\f{K}_{\rm in}=K_y\f{e}_y$ to $\f{K}_{\rm out}=-K_y\f{e}_y$ 
and thus they do not change the sign of $\dot E$ in Eq.~\eqref{transfer}.
\end{thebibliography}
\end{document}